\documentclass[aps, prd, amsmath, amssymb, preprintnumbers, groupedaddress, superscriptaddress, twocolumn, nofootinbib, 10pt]{revtex4-1}

\usepackage[hidelinks]{hyperref}

\makeatletter
\renewcommand{\p@subsection}{}
\renewcommand{\p@subsubsection}{}

\usepackage{
adjustbox,
amsthm,
booktabs,
mathtools,
tabularx,
tikz,
xcolor,
}

\usetikzlibrary{decorations.markings, calc,patterns}
\tikzstyle{stuff_fill_black}=[circle,draw,fill=black!]
\tikzstyle{stuff_fill_green}=[circle,draw,fill=green!]
\tikzstyle{stuff_fill_red}=[circle,draw,fill=red!40]
\tikzstyle{stuff_fill_blue}=[circle,draw,fill=cyan!70]
\tikzstyle{stuff_fill_connect}=[circle,draw,fill=orange!70]

\newcolumntype{C}{>{$}c<{$}}
\AtBeginDocument{
\heavyrulewidth=.08em
\lightrulewidth=.05em
\cmidrulewidth=.03em
\belowrulesep=.65ex
\belowbottomsep=0pt
\aboverulesep=.4ex
\abovetopsep=0pt
\cmidrulesep=\doublerulesep
\cmidrulekern=.5em
\defaultaddspace=.5em
}

\usepackage[LGR,T1]{fontenc}
\usepackage[utf8]{inputenc}
\usepackage[english]{babel}
\usepackage{cleveref}

\def\KB{{\overline{K}_{B_3}}}

\newtheoremstyle{break}  
  {\topsep}   
  {\topsep}   
  {}  
  {0pt}       
  {\bfseries} 
  {:}         
  {\newline}  
  {}          

\theoremstyle{break}
\newtheorem*{corollary}{Corollary}
\newtheorem*{prop}{Proposition}

\makeatletter
\let\@addpunct\@gobble
\makeatother
\newenvironment{myproof}[1][\bfseries \proofname]{%
  \begin{proof}[#1]$ $\par\nobreak\ignorespaces
}{%
  \end{proof}
}

\makeatletter
\DeclareOldFontCommand{\sc}{\normalfont\scshape}{\@nomath\sc}
\makeatother

\hypersetup{
    pdftitle={Statistics of Limit Root Bundles Relevant for Exact Matter Spectra of F-Theory MSSMs},
    pdfauthor={Martin Bies, Mirjam Cvetic, Muyang Liu},
    pdfsubject={Root bundles, Limit roots, Line bundles, Global sections, F-theory, MSSM, Quadrillion Standard Models, vector-like spectra, Statistics}
}

\begin{document}

\preprint{\texttt{UPR-1310-T}}
\title{Statistics of Limit Root Bundles\\
Relevant for Exact Matter Spectra of F-Theory MSSMs}
\author{Martin Bies}
\affiliation{Department of Mathematics, University of Pennsylvania, Philadelphia, PA 19104-6396, USA}
\affiliation{Department of Physics and Astronomy, University of Pennsylvania, Philadelphia, PA 19104-6396, USA}
\author{Mirjam Cveti\v{c}} \affiliation{Department of Physics and Astronomy, University of Pennsylvania, Philadelphia, PA 19104-6396, USA}
\affiliation{Department of Mathematics, University of Pennsylvania, Philadelphia, PA 19104-6396, USA}
\affiliation{Center for Applied Mathematics and Theoretical Physics, University of Maribor, Maribor, Slovenia}
\author{Muyang Liu} \affiliation{Department of Physics and Astronomy, University of Pennsylvania, Philadelphia, PA 19104-6396, USA}

\begin{abstract}
\noindent
In the largest, currently known, class of one Quadrillion globally consistent F-theory Standard Models with gauge coupling unification and no chiral exotics, the vector-like spectra are counted by cohomologies of root bundles. In this work, we apply a previously proposed method to identify toric base 3-folds, which are promising to establish F-theory Standard Models with exactly three quark-doublets and no vector-like exotics in this representation. The base spaces in question are obtained from triangulations of 708 polytopes. By studying root bundles on the quark doublet curve $C_{(\mathbf{3},\mathbf{2})_{1/6}}$ and employing well-known results about desingularizations of toric K3-surfaces, we derive a \emph{triangulation independent lower bound} $\check{N}_P^{(3)}$ for the number $N_P^{(3)}$ of root bundles on $C_{(\mathbf{3},\mathbf{2})_{1/6}}$ with exactly three sections. The ratio $\check{N}_P^{(3)} / N_P$, where $N_P$ is the total number of roots on $C_{(\mathbf{3},\mathbf{2})_{1/6}}$, is largest for base spaces associated with triangulations of the 8-th 3-dimensional polytope $\Delta^\circ_8$ in the Kreuzer-Skarke list. For each of these $\mathcal{O}( 10^{15} )$ 3-folds, we expect that many root bundles on $C_{(\mathbf{3},\mathbf{2})_{1/6}}$ are induced from F-theory gauge potentials and that at least every 3000th root on $C_{(\mathbf{3},\mathbf{2})_{1/6}}$ has exactly three global sections and thus no exotic vector-like quark-doublet modes.
\end{abstract}

\maketitle

\parskip 2pt plus 1pt minus 1pt
\interfootnotelinepenalty=10000



\section{Introduction}

Like no other framework for quantum gravity, string theory encodes the consistent coupling of gauge dynamics to gravity. Therefore, it is a leading candidate for a unified theory that accounts for all aspects of the observed low energy physics. Enormous efforts have been undertaken to demonstrate the particle spectrum of the Standard Model from string theory. The earliest studies focus on the $E_8 \times E_8$ heterotic string \cite{Candelas:1985en,Greene:1986ar,Braun:2005ux,Bouchard:2005ag,Bouchard:2006dn,Anderson:2009mh,Anderson:2011ns,Anderson:2012yf} and were later extended by intersecting branes models \cite{Berkooz:1996km,Aldazabal:2000dg,Aldazabal:2000cn,Ibanez:2001nd,Blumenhagen:2001te,Cvetic:2001tj,Cvetic:2001nr, Blumenhagen:2005mu}.

While these compactifications realize the gauge sector and chiral spectrum of the Standard Model, they are limited to the perturbative regime in the string coupling. Typically, they also suffer from vector-like exotics. The first globally consistent, perturbative MSSM constructions are \cite{Bouchard:2005ag,Bouchard:2006dn} (see \cite{Gomez:2005ii,Bouchard:2008bg} for more details).

In string compactifications, a significant amount of information is encoded in the geometry of the compactification space. A coherent approach to analyze the relations between geometry and physics is F-theory \cite{Vafa:1996xn,oai:arXiv.org:hep-th/9602114,oai:arXiv.org:hep-th/9603161}. It describes the gauge dynamics of 7-branes including their back-reactions \emph{to all orders in string coupling}. In 4-dimensional compactifications, this is achieved by encoding the back-reactions in the geometry of a singular elliptically fibered Calabi-Yau 4-fold $\pi \colon Y_4 \twoheadrightarrow B_3$. The global consistency conditions of the 4-dimensional physics are enforced by studying the geometry of $Y_4$, e.g., by a smooth, flat, crepant resolution $\widehat{Y}_4$.

The chiral spectrum of 4d $\mathcal{N} = 1$ F-theory compactifications is determined by a background $G_4$-flux. This flux is specified by the internal $C_3$ profile of the dual M-theory compactification via $G_4 = dC_3 \in H^{2,2} (\widehat{Y}_4)$, where $H^{2,2}(\widehat{Y}_4)$ is the middle vertical fourth cohomology of $\widehat{Y}_4$. The primary vertical subspace of $G_4$-configurations has been studied extensively \cite{oai:arXiv.org:1111.1232,Krause:2012yh,Braun:2013nqa,Cvetic:2013uta,Cvetic:2015txa,Lin:2015qsa,Lin:2016vus}. Applications to globally consistent chiral F-theory models \cite{Krause:2011xj,Cvetic:2015txa,Lin:2016vus,Cvetic:2018ryq} have lead to the discovery of the largest, currently-known, class of one Quadrillion globally consistent F-theory Standard Models (QSMs) with gauge coupling unification and no chiral exotics \cite{Cvetic:2019gnh}.

The massless vector-like spectrum depends not only on $G_4$, but also on the $C_3$-flat directions. The full gauge information is encoded in \emph{Deligne cohomology}. In \cite{Bies:2014sra,Bies:2017fam,Bies:2018uzw}, F-theory gauge potentials were parametrized by Chow classes, which in turn induce line bundles $L_{\mathbf{R}}$ on the matter curves $C_\mathbf{R} \subset B_3$. The (vector-like) zero modes are counted by the cohomologies of these line bundles.

In principle, this approach works for any compactification. Technical limitations arise in explicit geometries due to the intricate complex structure dependence of the cohomologies $h^i(C_{\mathbf{R}}, \, L_{\mathbf{R}})$ of the line bundles $L_{\mathbf{R}}$ on the matter curves $C_\mathbf{R}$. This dependence was investigated for special examples of F-theory compactifications in \cite{Bies:2020gvf}. A large data set was generated \cite{Database, ToricVarietiesProject} and investigated with data science techniques and completely understood by Brill-Noether theory \cite{Brill1874} (cf., \cite{Eisenbud1996, Watari:2016lft}).

For the QSMs \cite{Cvetic:2019gnh} another complication arises. As explained in \cite{bies2021root}, in these models the line bundles $L_{\mathbf{R}}$ are necessarily root bundles $P_{\mathbf{R}}$, which one may think of as generalizations of spin bundles. Just as spin bundles, there are typically $N_P( C_{\mathbf{R}} ) \gg 1$ root bundles on $C_{\mathbf{R}}$. Some of the $N_P( C_{\mathbf{R}} )$ roots stem from F-theory gauge backgrounds which induce the same chiral index but differ in their $C_3$-flat directions. An important task is to find the roots which are induced from F-theory gauge potentials and have cohomologies that define Minimal Supersymmetric Standard Models (MSSMs).

As a first step, a ``bottom-up'' analysis was conducted in \cite{bies2021root}. This work does not identify exactly which root bundles on $C_{\mathbf{R}}$ are induced from F-theory gauge potentials in the Deligne cohomology. Rather, a systematic study of the cohomologies of all admissable root bundles on $C_{\mathbf{R}}$ has been performed. Except for the Higgs curve, the prime interest are the \mbox{$N_P^{(3)} ( C_{\mathbf{R}} ) \leq N_P ( C_{\mathbf{R}} )$} roots with exactly three sections. By extending the results in \cite{2004math......4078C}, the authors formulated a technique to derive a lower bound $\check{N}_P^{(3)} ( C_{\mathbf{R}} )$ to $N_P^{(3)} ( C_{\mathbf{R}} )$.

The toric base spaces of the QSMs are obtained from triangulations of 708 polytopes in Kreuzer-Skarke list \cite{Kreuzer:1998vb}. The goal of this letter is to explain that \emph{$\check{N}_P^{(3)} ( C_{\mathbf{R}} )$ is independent of the triangulations}. We use this observation to identify promising toric 3-folds for F-Theory Standard Models without vector-like exotics on the quark-doublet curve $C_{(\mathbf{3},\mathbf{2})_{1/6}}$.

In \cref{subsec:Genesis} we recall the relation of the toric QSM base 3-folds and toric K3-surfaces. By studying limit roots on a nodal curve $C_{(\mathbf{3},\mathbf{2})_{1/6}}^\bullet$ and employing results of resolutions \cite{Batyrev:1994hm, Perevalov:1997vw, cox1999mirror, Rohsiepe:2004st,Braun:2017nhi}, we demonstrate in \cref{sec:TIndependence} that the derived lower bound $\check{N}_P^{(3)} ( C_{\mathbf{R}} )$ is \emph{independent of the triangulation}. We utilize the \texttt{Gap4}-package \emph{QSMExplorer} \cite{ToricVarietiesProject} in \cref{subsec:ScanOverQSM} to compute the ratio $\check{N}_P^{(3)} / N_P$ for several classes of toric QSM base 3-folds. We focus on bases, for which it can be expected that many root bundles on $C_{(\mathbf{3},\mathbf{2})_{1/6}}$ are ``top-down'' determined by gauge potentials of the F-theory compactification. This points us to the 3-folds associated with the $\mathcal{O}( 10^{15} )$ triangulations \cite{Halverson:2016tve} of the 8-th polytope $\Delta_8^\circ$ in the Kreuzer-Skarke list \cite{Kreuzer:1998vb}: At least every 3000-th root on $C_{(\mathbf{3},\mathbf{2})_{1/6}}$ has exactly three global sections and thus no vector-like exotics.

\section{Genesis of 3-fold bases} \label{subsec:Genesis}

Desingularizations of Calabi-Yau (CY) hypersurfaces in toric ambient space are studied in \cite{Batyrev:1994hm}. We focus on CY 2-folds, i.e. toric K3-surfaces. Those are associated to three-dimensional, reflexive lattice polytopes $\Delta \subset M_{\mathbb{R}}$ and their polar duals $\Delta^\circ \subset N_{\mathbb{R}}$, defined by $\left\langle \Delta, \Delta^\circ \right\rangle \geq -1$. Kreuzer and Skarke list all possible 3-dimensional polytopes \cite{Kreuzer:1998vb}. We consider the i-th polytope in the Kreuzer-Skarke list as subset of $N_{\mathbb{R}}$ and denote it by $\Delta_i^\circ$.

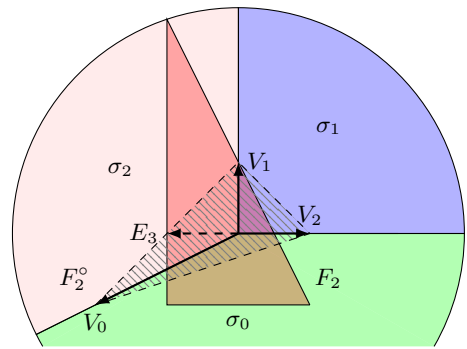
\begin{figure}[ht]
\begin{tikzpicture}
    \def\w{0.95};
    
    \draw (0,0) -- ({sqrt(5)*\w},0);
    \filldraw[fill opacity=0.3,fill=blue] (0,0) -- ({sqrt(10)*\w},0) arc (0:90:{sqrt(10)*\w}) -- cycle;
    \filldraw[fill opacity=0.3,fill=pink] (0,0) -- (0,{sqrt(10)*\w}) arc (90:206.565:{sqrt(10)*\w}) -- cycle;
    \filldraw[fill opacity=0.3,fill=green] (0,0) -- (-153.435:{sqrt(10)*\w}) arc (-153.435:-150:{sqrt(10)*\w});
    \filldraw[fill opacity=0.3,fill=green] (0,0) -- (0:{sqrt(10)*\w}) arc (0:-30:{sqrt(10)*\w});
    \fill[fill opacity=0.3,fill=green] (0,0) -- (-30:{sqrt(10)*\w}) -- (-150:{sqrt(10)*\w}) -- (0,0);
    
    \node at (45:{sqrt(5)*\w*0.8}) [above] {$\sigma_1$};
    \node at (157.5:{sqrt(5)*\w*0.8}) [above] {$\sigma_2$};
    \node at (0,{sqrt(5)*\w*(-0.65)}) [above] {$\sigma_0$};
    \node at (0.9,-0.6) [right] {$F_2$};
    \node at (-2.5,-0.6) [right] {$F_2^\circ$};
    
    \filldraw[fill opacity = 0.3, fill = red] (\w,-\w) -- (-\w,3*\w) -- (-\w,-\w) -- (\w,-\w);
    
    \draw[dashed,pattern=north west lines, pattern color=gray] (-153.435:{sqrt(5)*\w})--(0,\w)--(\w,0)--(-153.435:{sqrt(5)*\w});
    
    \node at (0,\w) [right] {$V_1$};
    \node at (\w,0) [above] {$V_2$};
    \node at (-153.435:{sqrt(5)*\w}) [below] {$V_0$};
    \node at (-\w,0) [left] {$E_3$};
    
    \draw [dashed, -latex, thick] (0,0) -- (-\w,0);
    \draw [-latex, thick] (0,0) -- (0,\w);
    \draw [-latex, thick] (0,0) -- (\w,0);
    \draw [-latex, thick] (0,0) -- (-153.435:{sqrt(5)*\w});
    
\end{tikzpicture}
\caption{MPCP of $F_2^\circ = \mathrm{Conv}(e_1,e_2,-2e_1-e_2 ) \subset N_{\mathbb{R}}$ refines normal fan $\Sigma_{F_2}$ of polytope $F_2 \subseteq M_{\mathbb{R}}$ by $E_3$.}
\label{fig:TwoFanF2}
\end{figure}

From a polytope $\Delta \subset M_{\mathbb{R}}$, one can build the normal fan $\Sigma_\Delta$. Its ray generators are the facet normals of $\Delta$ and the maximal cones are in one-to-one correspondence to the vertices of $\Delta$. We give a two-dimensional example in \cref{fig:TwoFanF2}. Neither the toric variety $X_{\Delta} \equiv X_{\Sigma_{\Delta}}$ nor the CY-hypersurface need be smooth. Resolutions of these CY-hypersurfaces were introduced in \cite{Batyrev:1994hm} as \emph{maximal projective crepant partial} (MPCP) desingularizations. Equivalently, \cite{cox1999mirror} refers to such desingularizations as maximal projective subdivisions of the normal fan.

To find MPCPs, we note that a refinement of the normal fan $\Sigma_{\Delta}$ by ray generators corresponding to lattice points of $\Delta^\circ$ is crepant. We can therefore consider refinements \mbox{$\Sigma( T ) \to \Sigma_{\Delta}$} where $\Sigma( T )$ is associated to a fine, regular, star triangulation (FRST) of the lattice polytope $\Delta^\circ$. We recall that \emph{star} means that every simplex in the triangulation contains the origin, \emph{fine} ensures that every lattice point of $\Delta^\circ$ is used as ray generator and that \emph{regular} implies that $X_{\Sigma(T)}$ is projective. Together, this implies that $\Sigma(T)$ defines a \emph{maximal} projective refinement of $\Sigma_{\Delta}$.

In our applications to toric K3-surfaces, $X_{\Sigma(T)}$ is guaranteed to be smooth. This is because a maximal projective subdivision of $\Sigma_{\Delta}$ then constructs a 3-dimensional Gorenstein orbifold with terminal singularities \cite{cox1999mirror} which must be smooth by proposition 11.4.19 in \cite{cox2011toric}. More details are provided in \cite{Batyrev:1994hm}. As an example, we indicate in \cref{fig:TwoFanF2} the two-dimensional polar dual polytope $F_2^\circ$ by dashed lines and the resulting refined fan $\Sigma(T)$.

\begin{figure}[ht]
    \centering
    \begin{tikzpicture}
    [x={(0.4497546cm, 0.559773cm)},
    y={(0.1306336cm, 0.681189cm)},
    z={(0.7360895cm, 0.0505690cm)},
    scale=1.80,
    back/.style={line width=1.0pt,opacity=0.50},
    edge/.style={color=brown, line width=2pt},
    facet/.style={fill=yellow!30,fill opacity=0.20},
    vertex/.style={inner sep=1.3pt,circle,fill=red!50,thick,anchor=base}]
    
    \coordinate (-1.0, -1.0, -1.0) at (-1.0, -1.0, -1.0);
    \coordinate (2.0, -1.0, 0.0) at (2.0, -1.0, 0.0);
    \coordinate (0.0, -1.0, -1.0) at (0.0, -1.0, -1.0);
    \coordinate (-1.0, 2.0, 0.0) at (-1.0, 2.0, 0.0);
    \coordinate (-1.0, -1.0, 3.0) at (-1.0, -1.0, 3.0);
    \coordinate (-1.0, 0.0, -1.0) at (-1.0, 0.0, -1.0);
    \coordinate (-1.0, -1.0, 0.0) at (-1.0, -1.0, 0.0);
    \coordinate (-1.0, -1.0, 1.0) at (-1.0, -1.0, 1.0);
    \coordinate (-1.0, -1.0, 2.0) at (-1.0, -1.0, 2.0);
    \coordinate (-1.0, 0.0, 0.0) at (-1.0, 0.0, 0.0);
    \coordinate (-1.0, 0.0, 1.0) at (-1.0, 0.0, 1.0);
    \coordinate (-1.0, 0.0, 2.0) at (-1.0, 0.0, 2.0);
    \coordinate (-1.0, 1.0, 0.0) at (-1.0, 1.0, 0.0);
    \coordinate (-1.0, 1.0, 1.0) at (-1.0, 1.0, 1.0);
    \coordinate (0.0, -1.0, 0.0) at (0.0, -1.0, 0.0);
    \coordinate (0.0, -1.0, 1.0) at (0.0, -1.0, 1.0);
    \coordinate (0.0, -1.0, 2.0) at (0.0, -1.0, 2.0);
    \coordinate (0.0, 0.0, 0.0) at (0.0, 0.0, 0.0);
    \coordinate (0.0, 0.0, 1.0) at (0.0, 0.0, 1.0);
    \coordinate (0.0, 1.0, 0.0) at (0.0, 1.0, 0.0);
    \coordinate (1.0, -1.0, 0.0) at (1.0, -1.0, 0.0);
    \coordinate (1.0, -1.0, 1.0) at (1.0, -1.0, 1.0);
    \coordinate (1.0, 0.0, 0.0) at (1.0, 0.0, 0.0);
    
    \draw[edge, black] (2.0, -1.0, 0.0) -- (0.0, -1.0, -1.0);
    \draw[edge, black] (-1.0, -1.0, -1.0) -- (0.0, -1.0, -1.0);
    \draw[edge, black] (2.0, -1.0, 0.0) -- (-1.0, 2.0, 0.0) -- (1,0,0) -- (0,1,0);
    \draw[edge, back] (-1.0, 2.0, 0.0) -- (-1.0, -1.0, 3.0) -- (-1,1,1) -- (-1,0,2);
    \draw[edge, black] (-1.0, -1.0, -1.0) -- (-1.0, -1.0, 3.0) -- (-1,-1,0) -- (-1,-1,1) -- (-1,-1,2);
    \draw[edge, black] (2.0, -1.0, 0.0) -- (-1.0, -1.0, 3.0) -- (1,-1,1) -- (0,-1,2);
    \draw[edge, black] (-1.0, -1.0, -1.0) -- (-1.0, 0.0, -1.0);
    \draw[edge, black] (0.0, -1.0, -1.0) -- (-1.0, 0.0, -1.0);
    \draw[edge, black] (-1.0, 2.0, 0.0) -- (-1.0, 0.0, -1.0);
    
    \fill[facet, fill opacity=0.30] (2.0, -1.0, 0.0) -- (-1.0, 2.0, 0.0) -- (-1.0, -1.0, 3.0) -- cycle {};
    \fill[facet, brown, fill opacity=0.40] (2.0, -1.0, 0.0) -- (0.0, -1.0, -1.0) -- (-1,0,-1) -- (-1.0, 2.0, 0.0) -- cycle {};
    \fill[facet, brown!80,fill opacity=0.30] (-1,-1,-1) -- (-1.0, -1.0, 3.0) -- (2.0, -1.0, 0.0) -- (0.0, -1.0, -1.0) -- cycle {};
    \fill[facet,  fill opacity=0.30] (-1,-1,-1) -- (-1.0, -1.0, 3.0) -- (-1.0, 2.0, 0.0) -- (-1.0, 0.0, -1.0) -- cycle {};
    \fill[facet, brown!80, fill opacity=0.30] (-1,-1,-1) -- (0.0, -1.0, -1.0) -- (-1.0, 0.0, -1.0) -- cycle {};
    
    \node[vertex] at  (-1.0, -1.0, -1.0) {};
    \node[vertex] at  (2.0, -1.0, 0.0) {};
    \node[vertex] at  (0.0, -1.0, -1.0) {};
    \node[vertex] at  (-1.0, 2.0, 0.0) {};
    \node[vertex] at  (-1.0, -1.0, 3.0) {};
    \node[vertex] at  (-1.0, 0.0, -1.0) {};
    \node[vertex] at  (-1.0, -1.0, 0.0) {};
    \node[vertex] at  (-1.0, -1.0, 1.0) {};
    \node[vertex] at  (-1.0, -1.0, 2.0) {};
    \node[vertex,gray] at  (-1.0, 0.0, 0.0) {};
    \node[vertex,gray] at  (-1.0, 0.0, 1.0) {};
    \node[vertex, fill opacity=0.30] at  (-1.0, 0.0, 2.0) {};
    \node[vertex,gray] at  (-1.0, 1.0, 0.0) {};
    \node[vertex, fill opacity=0.30] at  (-1.0, 1.0, 1.0) {};
    \node[vertex,gray, fill opacity=0.30] at  (0.0, -1.0, 0.0) {};
    \node[vertex,gray, fill opacity=0.30] at  (0.0, -1.0, 1.0) {};
    \node[vertex] at  (0.0, -1.0, 2.0) {};
    \node[vertex, magenta, fill opacity=0.60] at  (0.0, 0.0, 0.0) {};
    \node[vertex,gray] at  (0.0, 0.0, 1.0) {};
    \node[vertex,cyan] at  (0.0, 1.0, 0.0) {};
    \node[vertex, gray, fill opacity=0.30] at  (1.0, -1.0, 0.0) {};
    \node[vertex] at  (1.0, -1.0, 1.0) {};
    \node[vertex, cyan] at  (1.0, 0.0, 0.0) {};
    
    \end{tikzpicture}
    \begin{tikzpicture}
    [x={(0.449656cm, 0.377639cm)},
    y={(-0.77770cm, -0.358578cm)},
    z={(-0.106936cm, 0.633318cm)},
    scale=1.80,
    back/.style={line width=1.0pt,opacity=0.50},
    edge/.style={color=brown, line width=2pt},
    facet/.style={fill=yellow!30,fill opacity=0.20},
    vertex/.style={inner sep=1.3pt,circle,fill=red!50,thick,anchor=base}]
    
    \coordinate (-1.0, -1.0, -1.0) at (-1.0, -1.0, -1.0);
    \coordinate (-1.0, -1.0, 0.0) at (-1.0, -1.0, 0.0);
    \coordinate (-1.0, -1.0, 1.0) at (-1.0, -1.0, 1.0);
    \coordinate (-1.0, -1.0, 2.0) at (-1.0, -1.0, 2.0);
    \coordinate (0.0, 0.0, 1.0) at (0.0, 0.0, 1.0);
    \coordinate (0.0, 1.0, 0.0) at (0.0, 1.0, 0.0);
    \coordinate (1.0, 0.0, 0.0) at (1.0, 0.0, 0.0);
    \coordinate (0.0, 0.0, 0.0) at (0.0, 0.0, 0.0);
    
    \draw[edge, black] (-1.0, -1.0, -1.0) -- (-1.0, -1.0, 2.0) -- (-1,-1,0) -- (-1,-1,1);
    \draw[edge, black] (-1.0, -1.0, 2.0) -- (0.0, 1.0, 0.0);
    \draw[edge, black] (-1.0, -1.0, -1.0) -- (0,1,0);
    \draw[edge, black] (-1.0, -1.0, -1.0) -- (1.0, 0.0, 0.0);
    \draw[edge, black] (-1.0, -1.0, 2.0) -- (1.0, 0.0, 0.0);
    \draw[edge, back] (0.0, 1.0, 0.0) -- (1.0, 0.0, 0.0);
    \draw[edge, back] (-1.0, -1.0, 2.0) -- (0.0, 0.0, 1.0);
    \draw[edge, back] (0.0, 1.0, 0.0) -- (0.0, 0.0, 1.0);
    \draw[edge, back] (1.0, 0.0, 0.0) -- (0.0, 0.0, 1.0);
    
    \fill[facet, fill opacity=0.30] (0.0, 1.0, 0.0) -- (-1.0, -1.0, -1.0) -- (1.0, 0.0, 0.0) -- (0,0,1)-- cycle {};
    \fill[facet, fill opacity=0.30] (-1.0, -1.0, 2.0) -- (0.0, 1.0, 0.0) -- (0,0,1) -- cycle {};
    \fill[facet, brown, fill opacity=0.30] (-1.0, -1.0, -1.0) -- (-1.0, -1.0, 2.0) -- (1,0,0) -- cycle {};
    \fill[facet, fill opacity=0.30] (-1.0, -1.0, 2.0) -- (1.0, 0.0, 0.0) -- (0,0,1) -- cycle {};
    \fill[facet, fill opacity=0.30] (-1,-1,-1) -- (-1.0, -1.0, 2.0) -- (0.0, 1.0, 0.0) -- cycle {};
    \fill[facet, brown, fill opacity=0.30] (-1,-1,-1) -- (-1.0, -1.0, 2.0) -- (0.0, 1.0, 0.0) -- cycle {};
    
    \node[vertex] at (-1.0, -1.0, -1.0) {};
    \node[vertex, cyan] at (-1.0, -1.0, 0.0) {};
    \node[vertex, cyan] at (-1.0, -1.0, 1.0) {};
    \node[vertex] at (-1.0, -1.0, 2.0) {};
    \node[vertex, magenta, fill opacity=0.60] at (0.0, 0.0, 0.0) {};
    \node[vertex, fill opacity=0.30] at (0.0, 0.0, 1.0) {};
    \node[vertex] at (0.0, 1.0, 0.0) {};
    \node[vertex] at (1.0, 0.0, 0.0) {};
    
    \end{tikzpicture}
\caption{$\Delta_{52}^\circ \subset N_{\mathbb{R}}$ on the left and $\Delta_{52} \subseteq M_{\mathbb{R}}$ on the right \cite{Kreuzer:1998vb}. The magenta point is the origin. The generic K3-surface meets trivially with gray divisors, in an irreducible curve with the pinks and in finite families of $\mathbb{P}^1$s with the cyans.}
\label{fig:Delta13AndDeltaDual13}
\end{figure}
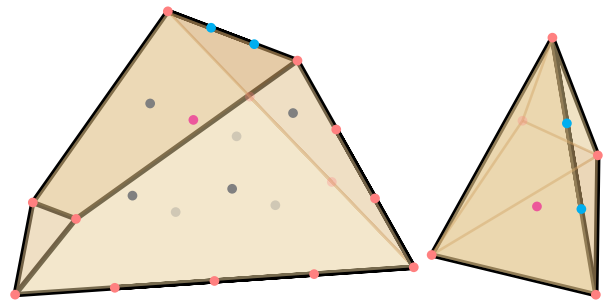

Of the 4319 polytopes in \cite{Kreuzer:1998vb}, 708 lead to toric 3-folds with $\overline{K}_{X_\Sigma( T )}^3 \in \{ 6, 10, 18, 30 \}$. Those are the base spaces for the Quadrillion F-theory Standard Models (QSMs) \cite{Cvetic:2019gnh}, in which the gauge divisors are K3-surfaces. This leads to gauge coupling unification. In the rest of this paper, we reserve the symbol $B_3 ( \Delta^\circ )$ for the family of all toric 3-folds obtained from FRSTs of the polytope $\Delta^\circ$. Our standing example in this letter will be the spaces build from the polytope $\Delta_{52}^\circ$ displayed in \cref{fig:Delta13AndDeltaDual13}.

\section{Triangulation independence} \label{sec:TIndependence}

\subsection{Dual graph} \label{subsec:DualGraph}

We will demonstrate that the dual graph of the nodal quark-doublet curve $C_{(\mathbf{3},\mathbf{2})_{1/6}}^\bullet$ introduced in \cite{bies2021root} is identical for all 3-folds $B_3( \Delta^\circ )$ obtained from FRSTs of $\Delta^\circ$. Hence, this dual graph only depends on $\Delta^\circ$.

The homogeneous coordinates of $X_{\Sigma(T)}$ correspond to the lattice points of $\Delta^\circ$. A coordinate associated to a facet interior point of $\Delta^\circ$ is denoted by $z_c$. For a lattice point in the interior of an edge $\Theta^\circ \subset \Delta^\circ$, two facets $F_1$, $F_2$ of $\Delta^\circ$ meet at $\Theta^\circ$. We notice that they are dual to vertices $m_1, m_2 \in \Delta$, and the dual edge $\Theta$ is the edge connecting $m_1$ and $m_2$. If $\Theta$ has interior points, we denote the homogeneous coordinate as $y_b$. All other coordinates are denoted as $x_a$. We mark these distinct types of lattice points in different colors in \cref{fig:Delta13AndDeltaDual13}. In particular, we write the nodal curve $C_{(\mathbf{3},\mathbf{2})_{1/6}}^\bullet$ introduced in \cite{bies2021root} as
\begin{align}
\adjustbox{max width=0.437\textwidth}{
$C_{(\mathbf{3},\mathbf{2})_{1/6}}^\bullet = \bigcup \limits_{a \in A}{V( x_a, s_9 )} \cup \bigcup \limits_{b \in B}{V( y_b, s_9 )} \cup \bigcup \limits_{c \in C}{V( z_c , s_9 )} \, ,$
} \label{equ:NodalCurve}
\end{align}
where $s_9$ is a generic section of $\overline{K}_{X_{\Sigma(T)}}$. The rational behind this classification is that will now explain that \emph{$V( x_a, s_9 )$ is irreducible, $V( y_b, s_9 )$ a finite collection of $\mathbb{P}^1$s and $V( z_c, s_9 ) = \emptyset$}.

We begin with $V( z_c, s_9 ) = \emptyset$, which was originally proven in \cite{Batyrev:1994hm,cox1999mirror} (see also \cite{Kreuzer:2006ax}). Since $X_{\Sigma( T )}$ is associated to a refinement of $\Sigma_{\Delta}$, there is a toric morphism $\varphi \colon X_{\Sigma( T )} \to X_\Delta$. By construction, this blow-down morphism is crepant and $V( z_c )$ is blown-down to a point, so that it does not intersect generic members of $| \overline{K}_{X_\Delta} |$. Since $\varphi$ is crepant and birational, $V( z_c )$ does therefore not intersect generic members of $\overline{K}_{\Sigma ( T )}$, i.e., $V( z_c , s_9 ) = \emptyset$. Hence, only the pink and cyan lattice points in \cref{fig:Delta13AndDeltaDual13} matter.

To see that $V( y_b, s_9 )$ is reducible, we compute its self-intersection in the K3-surface $V( s_9 )$. More generally, topological intersection numbers capture important properties of $C_{(\mathbf{3},\mathbf{2})_{1/6}}^\bullet$. For example, a curve component $C_i$ associated to $D_i \in \mathrm{Div}_T( X_{\Sigma( T )} )$ has \emph{arithmetic} genus $g( C_i )$ with $2 g(C_i) - 2 = D_i^2 \overline{K}_{X_{\Sigma(T)}}$. Similarly, the topological intersection of $C_i$ and $C_j$ is given by $D_i D_j \overline{K}_{X_{\Sigma(T)}}$. From the original work \cite{Perevalov:1997vw} (see also \cite{Rohsiepe:2004st}), it follows that these intersection numbers are counted by properties of $(\Delta^\circ, \Delta )$ and are thus independent of the FRST. Let us briefly restate this result.

\begin{prop}
Let $D_1, D_2 \in \mathrm{Div}_T( X_{\Sigma( T )} )$ be two distinct divisors corresponding to lattice points $v_1, v_2 \in \Delta^\circ$. If $v_1$, $v_2$ are not contained in an edge $\Theta^\circ \subset \Delta^\circ$, then $D_1 D_2 \overline{K}_{X_{\Sigma(T)}} = 0$. Otherwise, $D_1 D_2 \overline{K}_{X_{\Sigma(T)}} = 1 + l^\prime( \Theta )$, where $l^\prime( \Theta )$ is the number of interior lattice points on the dual edge $\Theta$.
\end{prop}

\begin{myproof}
Consider a triangulation $T$ of $\Delta^\circ$. Then the triple intersection among divisors $D_1$, $D_2$ and $\overline{K}_{X_{\Sigma(T)}}$ vanishes unless $v_1$, $v_2$ belong to two triangles in $T$, which we denote as $v_1 v_2 v_3$ and $v_1 v_2 v_4$.  It follows $D_1 D_2 D_i = 0$ if $i \notin \{ 1,2,3,4 \}$ and $D_1 D_2 D_3 = D_1 D_2 D_4 = 1$. Hence
\begin{align}
D_1 D_2 \overline{K}_{X_{\Sigma(T)}} = D_1 D_2 \left( D_1 + D_2 + D_3 + D_4 \right) \, . \label{equ:Intersection}
\end{align}
The affine span of $v_1$, $v_2$, $v_3$ contains a facet $F_3$ of $\Delta^\circ$. The dual of this facet is a vertex $m_3 \in \Delta$ with $\left\langle m_3, v_i \right\rangle = -1$. Let $N = \mathrm{rk} \left( \mathrm{Div}_T( X_{\Sigma(T)}) \right)$, then it holds $0 \sim \sum_{i = 1}^{N}{\left\langle m_3, v_i \right\rangle D_i}$ and thus
\begin{align}
D_1 + D_2 + D_3 \sim \left\langle m_3, v_4 \right\rangle D_4 + S \, ,
\end{align}
where $S$ satisfies $S D_1 D_2 = 0$. By substituting this back into \cref{equ:Intersection} we find $D_1 D_2 \overline{K}_{X_{\Sigma(T)}} = 1 + \left\langle m_3, v_4 \right\rangle$. If $v_1, v_2$ are not contained in an edge $\Theta^\circ \subset \Delta^\circ$, then $v_4 \in F_3$, $\left\langle m_3, v_4 \right\rangle = -1$ and $D_1 D_2 \overline{K}_{X_{\Sigma(T)}} = 0$.

Conversely, if $v_1, v_2 \in \Theta^\circ \subseteq \Delta^\circ$, then $v_1 v_2 v_4$ is contained in a facet $F_4 \neq F_3$ of $\Delta^\circ$ with dual vertex $m_4 \in \Delta$. The dual edge $\Theta$ from $m_3$ to $m_4$ only depends on $v_1$, $v_2$ but not the triangulation $T$. We now compare the number of interior lattice points $l^\prime( \Theta )$ on $\Theta$ to
\begin{align}
I_{12} = D_1 D_2 \overline{K}_{X_{\Sigma(T)}} = 1 + \left\langle m_3, v_4 \right\rangle \, .
\end{align}
$v_1$, $v_2$, $v_4$ generate $N_{\mathbb{R}}$. Therefore, $m \in M_{\mathbb{R}}$ is a lattice point iff $\left\langle m, v_1 \right\rangle$, $\left\langle m, v_2 \right\rangle$ and $\left\langle m, v_4 \right\rangle$ are integers. This shows that the lattice points on $\Theta$ are of the form
\begin{align}
m( k ) = m_3 + \left( \frac{1+k}{I_{12}} \right) \cdot ( m_4 - m_3 ) \, ,
\end{align}
where $k \in \{ -1, \dots, I_{12} \}$. Hence $l^\prime( \Theta ) = I_{12} - 1$.
\end{myproof}


We extend this to the arithmetic genera by restating another result from \cite{Perevalov:1997vw}.

\begin{corollary}
Let $D_1 \in \mathrm{Cl}( X_{\Sigma(T)} )$ be the divisor associated to the lattice point $v_1 \in \Delta^\circ$. Then $D_1^2 \overline{K}_{X_{\Sigma(T)}}$ is independent of triangulations of $\Delta^\circ$. Furthermore, if $v_1$ is an interior point of an edge $\Theta^\circ \subset \Delta^\circ$, then $D_1^2 \overline{K}_{X_{\Sigma(T)}} = - 2 - 2 l^\prime( \Theta )$.
\end{corollary}

\begin{myproof}
We consider a facet $F \subset \Delta^\circ$ with $v_1 \in F$. The dual vertex $m \in \Delta \in M_{\mathbb{R}}$ establishes $0 \sim \sum_{i = 1}^{N}{\left\langle m, v_i \right\rangle D_i}$ and hence $D_1^2 \overline{K}_{X_{\Sigma(T)}} = \sum_{i = 2}^{N}{\left\langle m, v_i \right\rangle \cdot D_1 D_i \overline{K}_{X_{\Sigma(T)}}}$, which is independent of FRSTs of $\Delta^\circ$ by the preceding proposition.

Next, assume that $v_1$ in an interior point of an edge $\Theta^\circ \subset \Delta^\circ$ and denote its neighbours along $\Theta^\circ$ by $v_2$ and $v_3$. The associated divisors $D_2$ and $D_3$ are the only divisors with non-zero $D_1 D_2 \overline{K}_{X_{\Sigma(T)}}$, $D_1 D_3 \overline{K}_{X_{\Sigma(T)}}$. $v_1$, $v_2$, $v_3$ are contained in a facet of $\Delta^\circ$, whose dual vertex $m \in \Delta$ establishes $D_1 \sim - D_2 - D_3 + S$ with $D_1 S \overline{K}_{X_{\Sigma(T)}} = 0$. Hence, as claimed, $D_1^2 \overline{K}_{X_{\Sigma(T)}} = - 2 - 2 l^\prime( \Theta )$.
\end{myproof}

$V( y_b, s_9 )$ corresponds to $v_b \in \Theta^\circ \subset \Delta^\circ$ with $l^\prime( \Theta ) > 0$. Hence $D_b^2 \cdot \overline{K}_{X_{\Sigma(T)}} = -2(l^\prime( \Theta ) + 1)< -2$ and $V( y_b, s_9 )$ is reducible into a collection of $l^\prime( \Theta ) + 1$ non-intersecting and smooth $\mathbb{P}^1$s \cite{Batyrev:1994hm, Perevalov:1997vw}.

Finally, let us turn to the components $V(x_a,s_9)$. A subset of these components is associated to lattice points $v_a \in \Theta^\circ$ such that $l^\prime( \Theta ) = 0$. By the previous result, these components are irreducible and smooth. The remaining $V( x_a, s_9 )$'s are associated to the vertices of $\Delta^\circ$. These components are smooth and irreducible by \cite{Batyrev:1994hm}.

\subsection{Computing the lower bound \texorpdfstring{$\check{N}_P^{(3)}$}{NP3C3216}} \label{subsec:TIndependentLowerBound}

We have established that in every space in $B_3( \Delta^\circ )$, $C^\bullet_{( \mathbf{3}, \mathbf{2} )_{1/6}}$ consists of the same components, with the same arithmetic genera and the same topological intersection numbers. Therefore, the dual graph $G( C^\bullet_{( \mathbf{3}, \mathbf{2} )_{1/6}} )$, in which components are vertices and intersections are edges, only depends on $\Delta^\circ$. We recall from \cite{bies2021root}, that on $C^\bullet_{( \mathbf{3}, \mathbf{2} )_{1/6}}$ we look for roots $P^\bullet_{(\mathbf{3},\mathbf{2})_{1/6}}$ subject to
\begin{align}
\adjustbox{max width=0.437\textwidth}{
$\left( P^\bullet_{(\mathbf{3},\mathbf{2})_{1/6}} \right)^{\otimes 2 \overline{K}_{X_{\Sigma(T)}}^3} = \left( \left. \overline{K}_{X_{\Sigma(T)}} \right|_{C^\bullet_{(\mathbf{3},\mathbf{2})_{1/6}}} \right)^{\otimes ( 6 + \overline{K}_{X_{\Sigma(T)}}^3 )} \, .$
}
\end{align}
Such roots are specified by weight assignments to $G( C^\bullet_{( \mathbf{3}, \mathbf{2} )_{1/6}} )$, constrained by $( 6 + \overline{K}_{X_{\Sigma(T)}}^3 )$-times the degree of $\left. \overline{K}_{X_{\Sigma(T)}} \right|_{C_i}$, where $C_i$ are the components of $C_{( \mathbf{3}, \mathbf{2} )_{1/6}}$. For $V( x_a, s_9 )$, $\left. \overline{K}_{X_{\Sigma(T)}} \right|_{C_i}$ has degree \mbox{$D_a \overline{K}_{X_{\Sigma(T)}}^2$} and for the irreducible components of $V( y_b, s_9 )$, this degree is $D_b \overline{K}_{X_{\Sigma(T)}}^2 / ( l^\prime( \Theta ) + 1 )$. Since $\overline{K}_{X_{\Sigma(T)}} = \sum_{i = 1}^{N}{D_i}$, we have $D \overline{K}_{X_{\Sigma(T)}}^2 = \sum_{i = 1}^{N}{D D_i \overline{K}_{X_{\Sigma(T)}}}$ and by the results of \cite{Perevalov:1997vw} restated in the previous section, these degrees are FRST-invariant. Similarly, $\overline{K}_{X_{\Sigma(T)}}^3$ is FRST-invariant. Consequently, the data that specifies the limit roots on $C^\bullet_{( \mathbf{3}, \mathbf{2} )_{1/6}}$ depends only on $\Delta^\circ$. By extending the techniques of \cite{bies2021root}, we can thus compute an FRST-invariant lower bound $\check{N}^{(3)}_P$ to the number of root bundles on $C_{( \mathbf{3}, \mathbf{2} )_{1/6}}$ with exactly three global sections.

We illustrate our computational strategy for $\check{N}^{(3)}_P$ with the polytope in \cref{fig:Delta13AndDeltaDual13} (see \cite{2004math......4078C, bies2021root} for more background). The MPCPs of $\Delta^\circ_{52}$ lead to toric 3-folds with $\KB^3 = 10$ and $h^{21} ( \widehat{Y}_4 ) = 7 > 6 = g \equiv g( C_{(\mathbf{3}, \mathbf{2})_{1/6}})$. The graph $G( C_{( \mathbf{3}, \mathbf{2} )_{1/6}} )$ looks as follows:
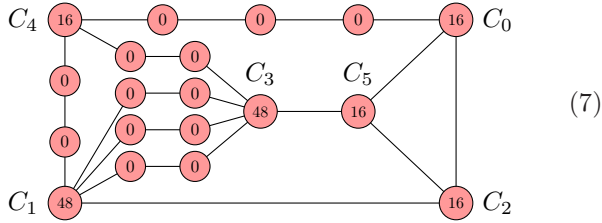
\begin{equation}
  \begin{tikzpicture}[scale=0.65, baseline=(current  bounding  box.center)]
      
      \def\s{2.0};
      \def\h{1.5};
      
      \path[-, every node/.append style={fill=white}] (-3*\s,2*\h) edge (-2.33*\s,1.5*\h);
      \path[-, every node/.append style={fill=white}] (-2.33*\s,1.5*\h) edge (-1.67*\s,1.5*\h);
      \path[-, every node/.append style={fill=white}] (-1.67*\s,1.5*\h) edge (-1*\s,0.75*\h);
      
      \path[-, every node/.append style={fill=white}] (-1*\s,0.75*\h) edge (0*\s,0.75*\h);
      \path[-, every node/.append style={fill=white}] (-\s,0.75*\h) edge (-1.67*\s,1.0*\h);
      \path[-, every node/.append style={fill=white}] (-\s,0.75*\h) edge (-1.67*\s,0.5*\h);
      \path[-, every node/.append style={fill=white}] (-\s,0.75*\h) edge (-1.67*\s,0.0*\h);
      \path[-, every node/.append style={fill=white}] (-3*\s,2*\h) edge (1*\s,2*\h);
      \path[-, every node/.append style={fill=white}] (-3*\s,-0.5*\h) edge (1*\s,-0.5*\h);
      \path[-, every node/.append style={fill=white}] (1*\s,2*\h) edge (1*\s,-0.5*\h);
      \path[-, every node/.append style={fill=white}] (1*\s,2*\h) edge (0*\s,0.75*\h);
      \path[-, every node/.append style={fill=white}] (1*\s,-0.5*\h) edge (0*\s,0.75*\h);
      \path[-, every node/.append style={fill=white}] (-3*\s,2*\h) edge (-3*\s,-0.5*\h);
      
      \path[-, every node/.append style={fill=white}] (-3*\s,-0.5*\h) edge (-2.33*\s,1.0*\h);
      \path[-, every node/.append style={fill=white}] (-3*\s,-0.5*\h) edge (-2.33*\s,0.5*\h);
      \path[-, every node/.append style={fill=white}] (-3*\s,-0.5*\h) edge (-2.33*\s,0.0*\h);
      \path[-, every node/.append style={fill=white}] (-2.33*\s,1.0*\h) edge (-1.67*\s,1.0*\h);
      \path[-, every node/.append style={fill=white}] (-2.33*\s,0.5*\h) edge (-1.67*\s,0.5*\h);
      \path[-, every node/.append style={fill=white}] (-2.33*\s,0.0*\h) edge (-1.67*\s,0.0*\h);
      
      \node at (-2.33*\s,1.5*\h) [stuff_fill_red, scale=0.65]{0};
      \node at (-1.67*\s,1.5*\h) [stuff_fill_red, scale=0.65]{0};
      \node at (-2.33*\s,1.0*\h) [stuff_fill_red, scale=0.65]{0};
      \node at (-2.33*\s,0.5*\h) [stuff_fill_red, scale=0.65]{0};
      \node at (-2.33*\s,0.0*\h) [stuff_fill_red, scale=0.65]{0};
      \node at (-1.67*\s,1.0*\h) [stuff_fill_red, scale=0.65]{0};
      \node at (-1.67*\s,0.5*\h) [stuff_fill_red, scale=0.65]{0};
      \node at (-1.67*\s,0.0*\h) [stuff_fill_red, scale=0.65]{0};
      \node at (-1*\s,0.75*\h) [stuff_fill_red, scale=0.6, label=above:$C_3$]{48};
      \node at (-3*\s,2*\h) [stuff_fill_red, scale=0.6, label=left:$C_4$]{16};
      \node at (-2.0*\s,2*\h) [stuff_fill_red, scale=0.65]{0};
      \node at (-1.0*\s,2*\h) [stuff_fill_red, scale=0.65]{0};
      \node at (0.0*\s,2*\h) [stuff_fill_red, scale=0.65]{0};
      \node at (1.0*\s,2*\h) [stuff_fill_red,  scale=0.6, label=right:$C_0$]{16};
      \node at (1.0*\s,-0.5*\h) [stuff_fill_red,  scale=0.6, label=right:$C_2$]{16};
      \node at (0*\s,0.75*\h) [stuff_fill_red,  scale=0.6, label=above:$C_5$]{16};
      \node at (-3*\s,1.1667*\h) [stuff_fill_red, scale=0.65]{0};
      \node at (-3*\s,0.3333*\h) [stuff_fill_red, scale=0.65]{0};
      \node at (-3*\s,-0.5*\h) [stuff_fill_red, scale=0.6, label=left:$C_1$]{48};
      
  \end{tikzpicture} \label{equ:DualGraph51}
  \end{equation}
The labels inside the nodes are the degree of $16\cdot \overline{K}_{C^\bullet_{( \mathbf{3}, \mathbf{2} )_{1/6}}}$ on the components $C_i$. To find the $( 2 \overline{K}_{X_{\Sigma(T)}}^3 )$-th roots we place weights $u_i, v_i \in \{ 1, 2, \dots, 2 \overline{K}_{X_{\Sigma(T)}}^3 - 1 \}$ along each edge in $G( C_{( \mathbf{3}, \mathbf{2} )_{1/6}} )$ subject to the following two rules (see \cite{2004math......4078C, bies2021root} for details):
\begin{enumerate}
    \item Along each edge, the sum of weights is $2 \overline{K}_{X_{\Sigma(T)}}^3$.
    \item At each node $C_i$, the sum of weights equals $( 6 + \overline{K}_{X_{\Sigma(T)}}^3 )$-times $D_i \overline{K}_{X_{\Sigma(T)}}^2$ modulo $2 \overline{K}_{X_{\Sigma(T)}}^3$.
\end{enumerate}
The number of possible weight assignments grows rapidly with the complexity of the dual graph. To speed up our counting procedure, we note that for counting $\check{N}^{(3)}_P$ it is possible to replace $G( C_{( \mathbf{3}, \mathbf{2} )^\bullet_{1/6}} )$ with a simplified graph. We remove components $C_i$ which are connected to exactly two other components and have $D_i \overline{K}_{X_{\Sigma(T)}}^2 = 0$. We are thus looking at transitions:
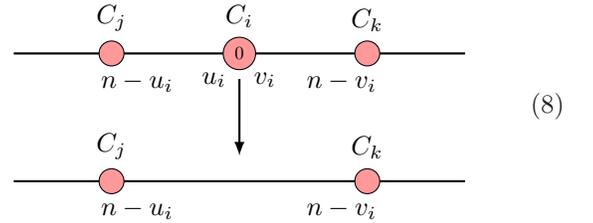
\begin{equation}
\begin{tikzpicture}[baseline=(current  bounding  box.center)]
    
    \def\w{3};
    \def\h{1.7};
    \def\o{1.5};
    
    \draw[thick] (-\w,0) -- (\w,0);
    \draw[thick] (-\w,-\h) -- (\w,-\h);
    
    \node at (-1*\h,0) [stuff_fill_red, label=above:$C_j$]{}; 
    \node at (0,0) [stuff_fill_red, scale=0.7, label=above:$C_i$]{0};
    \node at (1*\h,0) [stuff_fill_red, label=above:$C_k$]{}; 
    \node at (-0.8*\h,0) [label=below:$n - u_i$]{};
    \node at (0.8*\h,0) [label=below:$n - v_i$]{};
    \node at (0.2*\h,0) [label=below:$v_i$]{};
    \node at (-0.2*\h,0) [label=below:$u_i$]{};
    
    \node at (-1*\h,-\h) [stuff_fill_red, label=above:$C_j$]{}; 
    \node at (1*\h,-\h) [stuff_fill_red, label=above:$C_k$]{}; 
    \node at (-0.8*\h,-\h) [label=below:$n - u_i$]{};
    \node at (0.8*\h,-\h) [label=below:$n - v_i$]{};
    
    \draw[thick,-latex] (0,-0.2*\h)--(0,-0.8*\h);
    
\end{tikzpicture} \label{equ:ThreeNodes}
\end{equation}
To see that this does not change the lower bound $\check{N}^{(3)}_P$, let us focus on $n$-th roots. Then $1\leq u_i, v_i \leq n-1$. For given $u_i$ the weight $v_i$ is uniquely fixed as $v_i=n-u_i$. Since, $D_i \overline{K}_{X_{\Sigma(T)}}^2 = 0$, the resulting root on $C_i$ has degree $-1$ and supports no non-trivial sections (see \cite{2004math......4078C, bies2021root} for details). Conversely, given the diagram at the bottom, we can reconstruct the top-line by noting that $v_i = n - u_i$. For \cref{equ:DualGraph51}, this simplification leads to the graph:
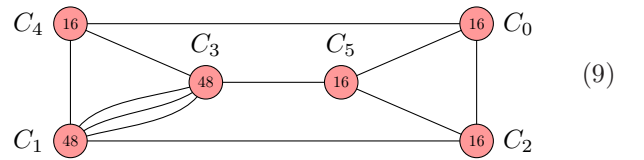
\begin{equation}
  \begin{tikzpicture}[scale=0.6, baseline=(current  bounding  box.center)]
      
      \def\s{1.5};
      \def\h{1.3};
      
      \path[-, every node/.append style={fill=white}] (-3*\s,\h) edge (3*\s,\h);
      \path[-, every node/.append style={fill=white}] (-3*\s,\h) edge (-3*\s,-\h);
      \path[-, every node/.append style={fill=white}] (3*\s,\h) edge (3*\s,-\h);
      \path[-, every node/.append style={fill=white}] (-1*\s,0) edge (1*\s,0);
      \path[-, every node/.append style={fill=white}] (-3*\s,-\h) edge (3*\s,-\h);
      \path[-, every node/.append style={fill=white}] (-3*\s,\h) edge (-1*\s,0);
      \path[-, every node/.append style={fill=white}] (3*\s,\h) edge (1*\s,0);
      \path[-, every node/.append style={fill=white}] (3*\s,-\h) edge (1*\s,0);
      
      \path[-, every node/.append style={fill=white}, out = 45, in = -135, looseness = 0.8] (-3*\s,-\h) edge (-1*\s,0);
      \path[-, every node/.append style={fill=white}, out = 20, in = -110, looseness = 0.8] (-3*\s,-\h) edge (-1*\s,0);
      \path[-, every node/.append style={fill=white}, out = 70, in = -160, looseness = 0.8] (-3*\s,-\h) edge (-1*\s,0);
      
      \node at (-3*\s,\h) [stuff_fill_red, scale=0.6, label=left:$C_4$]{16};
      \node at (-3*\s,-\h) [stuff_fill_red, scale=0.6, label=left:$C_1$]{48};
      \node at (-1*\s,0) [stuff_fill_red, scale=0.6, label=above:$C_3$]{48};
      \node at (3*\s,\h) [stuff_fill_red, scale=0.6, label=right:$C_0$]{16};
      \node at (3*\s,-\h) [stuff_fill_red, scale=0.6, label=right:$C_2$]{16};
      \node at (1*\s,0) [stuff_fill_red, scale=0.6, label=above:$C_5$]{16};
      
  \end{tikzpicture}
  \end{equation}
The algorithmic task of finding all weight assignments and counting the associated limit roots, can be conducted with the help of a computer implementation. The algorithms employed for this work are available in the \texttt{Gap4}-package \emph{QSMExplorer}, as part of the \emph{ToricVarieties$\_$project} \cite{ToricVarietiesProject}. On the computer \texttt{plesken.mathematik.uni-siegen.de}, our algorithm completes for $\Delta^\circ_{51}$ in roughly three minutes and finds $\check{N}^{(3)}_P = 34.980.351$. This number is to be compared to the total number of root bundles $N_P = 20^{12}$ on this $g = 6$ curve. Hence, at least every $1.2 \times 10^8$-th root on $C_{(\mathbf{3}, \mathbf{2})_{1/6}}$ has exactly three global sections.

\subsection{Towards favorable F-theory base spaces} \label{subsec:ScanOverQSM}

We extend this analysis to several classes $B_3( \Delta^\circ )$. Among the 708 polytopes, we focus on base spaces for which it can be expected that many roots stem from inequivalent F-theory gauge potentials in the Deligne cohomology $H^4_D( \widehat{Y}_4, \mathbb{Z}(2) )$, i.e., gauge potentials which induce the same chiral index but differ in their $C_3$-flat directions. In the 4-fold geometry $\widehat{Y}_4$, these $C_3$-flat directions are described by the intermediate Jacobian $J^2( \widehat{Y}_4)$. Since $h^{3,0}( \widehat{Y}_4 ) = 0$, we have (see \cite{Bies:2014sra} for more details)
\begin{align}
J^2( \widehat{Y}_4 ) = H^{2,1}( \widehat{Y}_4 ) / H^3( \widehat{Y}_4, \mathbb{Z} ) \, ,
\end{align}
and $\mathrm{dim}_{\mathbb{C}} ( J^2( \widehat{Y}_4 ) ) = h^{2,1}( \widehat{Y}_4 )$. In particular, if a gauge potential in $H^4_D( \widehat{Y}_4, \mathbb{Z}(2) )$ admits $( 2 \overline{K}_{X_{\Sigma(T)}}^3 )$-th roots, then it admits $( 2 \overline{K}_{X_{\Sigma(T)}}^3 )^{2 h^{2,1}( \widehat{Y}_4 )}$ roots. On the genus $g$ curve $C_{(\mathbf{3}, \mathbf{2})_{1/6}}$, we find $( 2 \overline{K}_{X_{\Sigma(T)}}^3 )^{2 g}$ roots. Therefore, a necessary condition for many roots on $C_{(\mathbf{3}, \mathbf{2})_{1/6}}$ to stem from F-theory gauge potentials is $h^{2,1}( \widehat{Y}_4 ) \geq g$:
\begin{align}
\begin{tabular}{cccc}
\toprule
$\KB^3$ & \# Polys & $h^{21}( \widehat{Y}_4 )$ & $g$ \\
\midrule
6 & 7 & $\{ 8,9,10,12,16 \}$  & 4 \\
10 & 54 & $\{ 2, 3, \dots, 11 \} \cup \{ 13 \}$ & 6 \\
18 & 373 & $\{ 0, 1, \dots, 12 \}$ & 10 \\
30 & 274 & $\{ 0, 1, \dots, 9 \}$ & 16 \\
\bottomrule
\end{tabular}
\end{align}
All 7 polytopes with $\KB^3 = 6$ satisfy this necessary condition. Their triangulations give at least $50\%$ of the QSM 3-fold base spaces \cite{Halverson:2016tve}. In addition, 27 polytopes with $\KB^3 = 10$ and three with $\KB^3 = 18$ have this property. For the $\KB^3 = 10$ polytope $\Delta_{14}^\circ$ and the three $\KB^3 = 18$ polytopes $\Delta^\circ_{72}$, $\Delta^\circ_{229}$ and $\Delta^\circ_{527}$, the quark-doublet curve has a component with genus larger than one. Hence, for these space the counting procedure introduced in \cite{bies2021root} does not apply. However, for the remaining 33 polytopes, our computer implementation finds $\check{N}^{(3)}_P( C_{(\mathbf{3}, \mathbf{2})_{1/6}} )$ within a few minutes. For the $\KB^3 = 6$ bases, we have $N_P( C_{(\mathbf{3}, \mathbf{2})_{1/6}} ) = 12^8$ and for those with $\KB^3 = 10$ we have $N_P( C_{(\mathbf{3}, \mathbf{2})_{1/6}} ) = 20^{12}$. The computed lower bounds are listed in \cref{table:Results}.

\begin{table}[tb]
\centering
\begin{tabular}{ccc||ccc}
\toprule
\multicolumn{6}{c}{$\KB^3 = 6$}\\
\midrule
& $\check{N}^{(3)}_P$ & $N_P / \check{N}^{(3)}_P$ & & $\check{N}^{(3)}_P$ & $N_P / \check{N}^{(3)}_P$ \\
\midrule
$\Delta^\circ_{8}$ & $142560$ & $3.0 \cdot 10^3$ & $\Delta^\circ_{130}$ & $8910$ & $4.8 \cdot 10^4$ \\
$\Delta^\circ_{4}$ & $11110$ & $3.8 \cdot 10^4$ & $\Delta^\circ_{136}$ & $8910$ & $4.8 \cdot 10^4$\\
$\Delta^\circ_{134}$ & $10100$ & $4.3 \cdot 10^4$ & $\Delta^\circ_{236}$ & $8910$ & $4.8 \cdot 10^4$ \\
$\Delta^\circ_{128}$ & $8910$ & $4.8 \cdot 10^4$ \\
\midrule \midrule
\multicolumn{6}{c}{$\KB^3 = 10$} \\
\midrule
& $\check{N}^{(3)}_P$ & $N_P / \check{N}^{(3)}_P$ & & $\check{N}^{(3)}_P$ & $N_P / \check{N}^{(3)}_P$\\
\midrule
$\Delta^\circ_{88}$ & $781.680.888$ & $5.2 \cdot 10^6$ & $\Delta^\circ_{762}$ & $32.858.151$ & $1.2 \cdot 10^8$ \\
$\Delta^\circ_{110}$ & $738.662.983$ & $5.5 \cdot 10^6$ & $\Delta^\circ_{417}$ & $32.857.596$ & $1.2 \cdot 10^8$ \\
$\Delta^\circ_{272}$ & $736.011.640$ & $5.6 \cdot 10^6$ & $\Delta^\circ_{838}$ & $32.845.047$ & $1.2 \cdot 10^8$\\
$\Delta^\circ_{274}$ & $736.011.640$ & $5.6 \cdot 10^6$ & $\Delta^\circ_{782}$ & $32.844.379$ & $1.2 \cdot 10^8$ \\
$\Delta^\circ_{387}$ & $733.798.30$ & $5.6 \cdot 10^6$ & $\Delta^\circ_{377}$ & $30.846.440$ & $1.3 \cdot 10^8$ \\
\midrule
$\Delta^\circ_{798}$ & $690.950.608$ & $5.9 \cdot 10^6$ & $\Delta^\circ_{499}$ & $30.846.440$ & $1.3 \cdot 10^8$ \\
$\Delta^\circ_{808}$ & $690.950.608$ & $5.9 \cdot 10^6$ & $\Delta^\circ_{503}$ & $30.846.440$ & $1.3 \cdot 10^8$ \\
$\Delta^\circ_{810}$ & $690.950.608$ & $5.9 \cdot 10^6$ & $\Delta^\circ_{1348}$ & $30.845.702$ & $1.3 \cdot 10^8$ \\
$\Delta^\circ_{812}$ & $690.950.608$ & $5.9 \cdot 10^6$ & $\Delta^\circ_{882}$ & $30.840.098$ & $1.3 \cdot 10^8$ \\
$\Delta^\circ_{254}$ & $35.004.914$ & $1.2 \cdot 10^8$ & $\Delta^\circ_{1340}$ & $28.954.543$ & $1.4 \cdot 10^8$ \\
\midrule
$\Delta^\circ_{52}$ & $34.980.351$ & $1.2 \cdot 10^8$ & $\Delta^\circ_{1879}$ & $28.950.852$ & $1.4 \cdot 10^8$ \\
$\Delta^\circ_{302}$ & $34.908.682$ & $1.2 \cdot 10^8$ & $\Delta^\circ_{1384}$ & $27.178.020$ & $1.5 \cdot 10^8$ \\
$\Delta^\circ_{786}$ & $32.860.461$ & $1.2 \cdot 10^8$ & $\Delta^\circ_{856}$ & $22.807.749$ & $1.8 \cdot 10^8$ \\
\bottomrule
\end{tabular} \caption{$\check{N}^{(3)}_P$ for 33 QSM polytopes with $h^{21}( \widehat{Y}_4 ) \geq g$.}
\label{table:Results}
\end{table}

Among these 33 polytopes, the ratio $\check{N}^{(3)}_P / N_P$ is largest for $\Delta^\circ_{8}$. In addition, within the QSMs, base spaces obtained from FRSTs of $\Delta^\circ_{8}$ have the maximal $16 = h^{21} ( \widehat{Y}_4 )$ and the minimal $g = 4$. In this sense, they most positively satisfy the necessary condition for a top-down origin of at least some of the root bundles. In this sense, these $\mathcal{O}( 10^{15} )$ toric base 3-folds \cite{Halverson:2016tve} are currently the most promising candidates to establish an F-theory Standard Model with exactly three quark-doublets and no vector-like exotics in this representation.

\section{Discussion and Outlook}

A construction of one Quadrillion globally consistent F-theory Standard Models (QSMs) with gauge coupling unification and no chiral exotics was presented in \cite{Cvetic:2019gnh}. In this work, we apply the techniques introduced in \cite{bies2021root} systematically to the toric QSM base 3-folds. Our goal is to identify toric base spaces, which are promising candidates to establish F-theory Standard Models with exactly three quark-doublets and no vector-like exotics in this representation.

We recall that vector-like spectra are counted by cohomologies of line bundles $L_{\mathbf{R}}$ on the matter curves $C_{\mathbf{R}}$. In \cite{bies2021root}, it was argued that these bundles must necessarily be root bundles. For instance, on the quark-doublet curve $C_{(\mathbf{3},\mathbf{2})_{1/6}}$ we consider line bundles $P_{(\mathbf{3},\mathbf{2})_{1/6}}$ which solve
\begin{align}
P_{(\mathbf{3},\mathbf{2})_{1/6}}^{\otimes 2 \KB^3} = K_{(\mathbf{3},\mathbf{2})_{1/6}}^{\otimes \left( 6 + \KB^3 \right)} \, , \label{equ:RootBundleConstraint}
\end{align}
where $\KB^3$ is the triple intersection number of the anticanonical class of the 3-fold $B_3$. This constraint has $N_P( C_{(\mathbf{3},\mathbf{2})_{1/6}} )= ( 2 \KB^3 )^{2g}$ solutions, where $g$ is the genus of $C_{(\mathbf{3},\mathbf{2})_{1/6}}$. In every QSM vacuum, the zero mode spectrum is counted by the cohomologies of one of these solutions. It is currently not known exactly which roots stem from F-theory gauge potentials in the Deligne cohomology $H^4_D( \widehat{Y}_4, \mathbb{Z}(2) )$ of the elliptic 4-fold $\widehat{Y}_4$.

In this work, we did not attempt to give a detailed answer to this question. Rather, we focused on base spaces for which it can be expected that many roots are induced from inequivalent F-theory gauge potentials, i.e., gauge potentials which induce the same chiral index but differ in their $C_3$-flat directions. In the 4-fold geometry $\widehat{Y}_4$, these $C_3$-flat directions are described by the intermediate Jacobian $J^2( \widehat{Y}_4)$. Since $h^{3,0}( \widehat{Y}_4 ) = 0$, it holds $J^2( \widehat{Y}_4 ) = H^{2,1}( \widehat{Y}_4 ) / H^3( \widehat{Y}_4, \mathbb{Z} )$ (see \cite{Bies:2014sra} for details) and $\mathrm{dim}_{\mathbb{C}} ( J^2( \widehat{Y}_4 ) ) = h^{2,1}( \widehat{Y}_4 )$. If a gauge potential in $H^4_D( \widehat{Y}_4, \mathbb{Z}(2) )$ admits $( 2 \KB^3 )$-th roots, then it admits $( 2 \KB^3 )^{2 h^{2,1}( \widehat{Y}_4 )}$ roots. On $C_{(\mathbf{3}, \mathbf{2})_{1/6}}$, we find $( 2 \KB )^{2 g}$ roots. Therefore, a necessary condition for many roots on $C_{(\mathbf{3}, \mathbf{2})_{1/6}}$ to stem from F-theory gauge potentials is $h^{2,1}( \widehat{Y}_4 ) \geq g$.

The QSM toric base 3-folds are obtained from fine, regular, star triangulations (FRSTs) of 708 3-dimensional, reflexive, lattice polytopes \cite{Cvetic:2019gnh}. In these spaces, the gauge divisors are K3-surfaces. This leads to gauge coupling unification and emphasizes the physical significance of $\KB$. Only 37 of the QSM polytopes lead to spaces with $h^{2,1}( \widehat{Y}_4 ) \geq g$. Still, the triangulations of these 37 polytopes provide the majority of the $\mathcal{O}( 10^{15} )$ toric QSM base 3-folds \cite{Halverson:2016tve,Cvetic:2019gnh}.

The natural next step is to count the number $N^{(3)}_P$ of roots which solve \cref{equ:RootBundleConstraint} and admit exactly three global sections, thus ensuring no vector-like exotics on the quark-doublet curve. In following \cite{bies2021root}, we achieve this by studying limit roots on a nodal curve $C^\bullet_{(\mathbf{3},\mathbf{2})_{1/6}}$ introduced in \cite{bies2021root}, which establishes a lower bound $\check{N}^{(3)}_P \leq N^{(3)}_P$. Crucially, we argue that \emph{$\check{N}^{(3)}_P$ is identical for all spaces $B_3( \Delta^\circ )$ obtained from FRSTs of $\Delta^\circ$, that is $\check{N}^{(3)}_P$ depends only on $\Delta^\circ$ and not the FRSTs.}

We establish this result by arguing that the data, which specifies the limit roots, is identical for all spaces in $B_3( \Delta^\circ )$. This in turn follows by noting that the QSM base spaces are obtained from desingularizations of toric K3-hypersurfaces. Since the nodal curve $C^\bullet_{(\mathbf{3},\mathbf{2})_{1/6}}$ is closely related to the Picard lattice of the resulting smooth, toric K3-surface, we could employ powerful and well-known results about such desingularizations \cite{Batyrev:1994hm, Perevalov:1997vw, cox1999mirror, Rohsiepe:2004st} (see also \cite{Braun:2017nhi} for recent work on related topics), and thereby establish the claim. Explicitly, this reduces to the FRST-invariance of topological triple-intersection numbers, which are related to FRST-independent counts of lattice points in the polytope $\Delta$ \cite{Perevalov:1997vw}.

Among the 37 polytopes with $h^{21}( \widehat{Y}_4 ) \geq g$, there are four polytopes for which $C_{(\mathbf{3},\mathbf{2})_{1/6}}$ has components with genus larger than one, so that the techniques introduced in \cite{bies2021root} cannot be applied. For the remaining 33 polytopes, we list the lower bounds in \cref{table:Results}. These counts were determined with the \texttt{Gap4} package \emph{QSMExplorer}, which is part of the \emph{ToricVarieties$\_$project} \cite{ToricVarietiesProject}. We have optimized the input for this algorithm by simplifying the dual graph of $C^\bullet_{(\mathbf{3},\mathbf{2})_{1/6}}$. For one polytope and a personal computer, we expect runtimes from a few seconds to around 10 minutes for the lower bounds in \cref{table:Results}.

Surprisingly, the simplifications of the dual graph of $C^\bullet_{(\mathbf{3},\mathbf{2})_{1/6}}$ lead to very similar graphs for distinct polytopes, and at times even identical lower bounds. For example, this applies to $\Delta^\circ_{128}$, $\Delta^\circ_{130}$, $\Delta^\circ_{136}$ and $\Delta^\circ_{236}$, for which we find $\check{N}^{(3)}_P = 8910$. We reserve a more detailed study of this phenomenon for future work.

Finally, we can read-off from \cref{table:Results} the spaces for which $\check{N}^{(3)}_P /N_P$ and $h^{21}( \widehat{Y}_4 ) / g( C_{(\mathbf{3}, \mathbf{2})_{1/6}} )$ are largest. This points us to $B_3( \Delta^\circ_8 )$. We can expect that at least every 3000-th root on $C_{(\mathbf{3},\mathbf{2})_{1/6}}$ has exactly three global sections. Furthermore, these spaces satisfy
\begin{align}
16 = h^{21}( \widehat{Y}_4 ) \geq g \equiv g( C_{(\mathbf{3},\mathbf{2})_{1/6}} ) = 4 \, ,
\end{align}
which is the largest, respective smallest possible value for all QSM spaces. Therefore, the triangulations of $\Delta^\circ_8$ lead to $\mathcal{O}( 10^{15} )$ \cite{Halverson:2016tve} promising toric base 3-folds for F-theory Standard Models with exactly three quark-doublets and no vector-like exotics in this representation.

The study of root bundles on the matter curves $C_{(\mathbf{1},\mathbf{1})_{1}}$ and $C_{(\mathbf{3},\mathbf{1})_{-2/3}}$ is identical to the presented study of roots on $C_{(\mathbf{3},\mathbf{2})_{1/6}}$. The matter curve $C_{(\mathbf{3},\mathbf{1})_{1/3}}$ is more complicated due to its higher genus, but can at least in principle be treated analogously. The real challenge however, is to establish one vector-like pair on the Higgs curve $C_{(\mathbf{1},\mathbf{2})_{-1/2}}$ and to investigate the ``top-down'' origin of the root bundles from F-theory gauge potentials. It can be anticipated that a detailed study of these questions will shed more light on the structure and construction of F-theory MSSMs. We hope to return to this thrilling and challenging task in the near future.

\begin{acknowledgments}

We are grateful to Ron Donagi and Marielle Ong for past collaboration, insightful discussions and ongoing work on vector-like spectra on the Higgs curve. The reliable computations of \texttt{plesken.mathematik.uni-siegen.de} are truly appreciated. We thank  Kamal Saleh and Jiahua Tian for valuable discussions. M.B is partially supported by NSF grant DMS 201673 and by the Simons Foundation Collaboration grant \#390287 on ``Homological Mirror Symmetry''. The work of M.C.~and M.L.~ is supported by DOE Award DE-SC013528Y. M.B.~and M.C.~further acknowledge support by the Simons Foundation Collaboration grant \#724069 on ``Special Holonomy in Geometry, Analysis and Physics''. M.C.~thanks the Slovenian Research Agency \mbox{No.~P1-0306} and the Fay R.~and Eugene L.~Langberg Chair for their support.
\end{acknowledgments}


\bibliography{paper}

\providecommand{\href}[2]{#2}\begingroup\raggedright\begin{thebibliography}{10}

\bibitem{Candelas:1985en}
P.~Candelas, G.T.~Horowitz, A.~Strominger and E.~Witten, \emph{{Vacuum
  Configurations for Superstrings}},
  \href{https://doi.org/10.1016/0550-3213(85)90602-9}{\emph{Nucl. Phys.}
  {\bfseries B258} (1985) 46}.

\bibitem{Greene:1986ar}
B.R.~Greene, K.H.~Kirklin, P.J.~Miron and G.G.~Ross, \emph{{A Superstring
  Inspired Standard Model}},
  \href{https://doi.org/10.1016/0370-2693(86)90137-1}{\emph{Phys. Lett. B}
  {\bfseries 180} (1986) 69}.

\bibitem{Braun:2005ux}
V.~Braun, Y.-H.~He, B.A.~Ovrut and T.~Pantev, \emph{{A Heterotic standard
  model}}, \href{https://doi.org/10.1016/j.physletb.2005.05.007}{\emph{Phys.
  Lett. B} {\bfseries 618} (2005) 252}
  [\href{https://arxiv.org/abs/hep-th/0501070}{{\ttfamily hep-th/0501070}}].

\bibitem{Bouchard:2005ag}
V.~Bouchard and R.~Donagi, \emph{{An SU(5) heterotic standard model}},
  \href{https://doi.org/10.1016/j.physletb.2005.12.042}{\emph{Phys. Lett.}
  {\bfseries B633} (2006) 783}
  [\href{https://arxiv.org/abs/hep-th/0512149}{{\ttfamily hep-th/0512149}}].

\bibitem{Bouchard:2006dn}
V.~Bouchard, M.~Cveti{\v c} and R.~Donagi, \emph{{Tri-linear couplings in an
  heterotic minimal supersymmetric standard model}},
  \href{https://doi.org/10.1016/j.nuclphysb.2006.03.032}{\emph{Nucl. Phys. B}
  {\bfseries 745} (2006) 62}
  [\href{https://arxiv.org/abs/hep-th/0602096}{{\ttfamily hep-th/0602096}}].

\bibitem{Anderson:2009mh}
L.B.~Anderson, J.~Gray, Y.-H.~He and A.~Lukas, \emph{{Exploring Positive Monad
  Bundles And A New Heterotic Standard Model}},
  \href{https://doi.org/10.1007/JHEP02(2010)054}{\emph{JHEP} {\bfseries 02}
  (2010) 054} [\href{https://arxiv.org/abs/0911.1569}{{\ttfamily 0911.1569}}].

\bibitem{Anderson:2011ns}
L.B.~Anderson, J.~Gray, A.~Lukas and E.~Palti, \emph{{Two Hundred Heterotic
  Standard Models on Smooth Calabi-Yau Threefolds}},
  \href{https://doi.org/10.1103/PhysRevD.84.106005}{\emph{Phys. Rev.}
  {\bfseries D84} (2011) 106005}
  [\href{https://arxiv.org/abs/1106.4804}{{\ttfamily 1106.4804}}].

\bibitem{Anderson:2012yf}
L.B.~Anderson, J.~Gray, A.~Lukas and E.~Palti, \emph{{Heterotic Line Bundle
  Standard Models}}, \href{https://doi.org/10.1007/JHEP06(2012)113}{\emph{JHEP}
  {\bfseries 06} (2012) 113} [\href{https://arxiv.org/abs/1202.1757}{{\ttfamily
  1202.1757}}].

\bibitem{Berkooz:1996km}
M.~Berkooz, M.R.~Douglas and R.G.~Leigh, \emph{{Branes intersecting at
  angles}}, \href{https://doi.org/10.1016/S0550-3213(96)00452-X}{\emph{Nucl.
  Phys.} {\bfseries B480} (1996) 265}
  [\href{https://arxiv.org/abs/hep-th/9606139}{{\ttfamily hep-th/9606139}}].

\bibitem{Aldazabal:2000dg}
G.~Aldazabal, S.~Franco, L.E.~Ibanez, R.~Rabadan and A.M.~Uranga, \emph{{D = 4
  chiral string compactifications from intersecting branes}},
  \href{https://doi.org/10.1063/1.1376157}{\emph{J. Math. Phys.} {\bfseries 42}
  (2001) 3103} [\href{https://arxiv.org/abs/hep-th/0011073}{{\ttfamily
  hep-th/0011073}}].

\bibitem{Aldazabal:2000cn}
G.~Aldazabal, S.~Franco, L.E.~Ibanez, R.~Rabadan and A.M.~Uranga,
  \emph{{Intersecting brane worlds}},
  \href{https://doi.org/10.1088/1126-6708/2001/02/047}{\emph{JHEP} {\bfseries
  02} (2001) 047} [\href{https://arxiv.org/abs/hep-ph/0011132}{{\ttfamily
  hep-ph/0011132}}].

\bibitem{Ibanez:2001nd}
L.E.~Ibanez, F.~Marchesano and R.~Rabadan, \emph{{Getting just the standard
  model at intersecting branes}},
  \href{https://doi.org/10.1088/1126-6708/2001/11/002}{\emph{JHEP} {\bfseries
  11} (2001) 002} [\href{https://arxiv.org/abs/hep-th/0105155}{{\ttfamily
  hep-th/0105155}}].

\bibitem{Blumenhagen:2001te}
R.~Blumenhagen, B.~Kors, D.~L{\"u}st and T.~Ott, \emph{{The standard model from
  stable intersecting brane world orbifolds}},
  \href{https://doi.org/10.1016/S0550-3213(01)00423-0}{\emph{Nucl. Phys.}
  {\bfseries B616} (2001) 3}
  [\href{https://arxiv.org/abs/hep-th/0107138}{{\ttfamily hep-th/0107138}}].

\bibitem{Cvetic:2001tj}
M.~Cveti{\v c}, G.~Shiu and A.M.~Uranga, \emph{{Three family supersymmetric
  standard - like models from intersecting brane worlds}},
  \href{https://doi.org/10.1103/PhysRevLett.87.201801}{\emph{Phys. Rev. Lett.}
  {\bfseries 87} (2001) 201801}
  [\href{https://arxiv.org/abs/hep-th/0107143}{{\ttfamily hep-th/0107143}}].

\bibitem{Cvetic:2001nr}
M.~Cveti{\v c}, G.~Shiu and A.M.~Uranga, \emph{{Chiral four-dimensional N=1
  supersymmetric type 2A orientifolds from intersecting D6 branes}},
  \href{https://doi.org/10.1016/S0550-3213(01)00427-8}{\emph{Nucl. Phys.}
  {\bfseries B615} (2001) 3}
  [\href{https://arxiv.org/abs/hep-th/0107166}{{\ttfamily hep-th/0107166}}].

\bibitem{Blumenhagen:2005mu}
R.~Blumenhagen, M.~Cveti{\v c}, P.~Langacker and G.~Shiu, \emph{{Toward
  realistic intersecting D-brane models}},
  \href{https://doi.org/10.1146/annurev.nucl.55.090704.151541}{\emph{Ann. Rev.
  Nucl. Part. Sci.} {\bfseries 55} (2005) 71}
  [\href{https://arxiv.org/abs/hep-th/0502005}{{\ttfamily hep-th/0502005}}].

\bibitem{Gomez:2005ii}
T.L.~Gomez, S.~Lukic and I.~Sols, \emph{{Constraining the Kahler moduli in the
  heterotic standard model}},
  \href{https://doi.org/10.1007/s00220-007-0338-8}{\emph{Commun. Math. Phys.}
  {\bfseries 276} (2007) 1}
  [\href{https://arxiv.org/abs/hep-th/0512205}{{\ttfamily hep-th/0512205}}].

\bibitem{Bouchard:2008bg}
V.~Bouchard and R.~Donagi, \emph{{On heterotic model constraints}},
  \href{https://doi.org/10.1088/1126-6708/2008/08/060}{\emph{JHEP} {\bfseries
  08} (2008) 060} [\href{https://arxiv.org/abs/0804.2096}{{\ttfamily
  0804.2096}}].

\bibitem{Vafa:1996xn}
C.~Vafa, \emph{{Evidence for F theory}},
  \href{https://doi.org/10.1016/0550-3213(96)00172-1}{\emph{Nucl. Phys.}
  {\bfseries B469} (1996) 403}
  [\href{https://arxiv.org/abs/hep-th/9602022}{{\ttfamily hep-th/9602022}}].

\bibitem{oai:arXiv.org:hep-th/9602114}
D.R.~Morrison and C.~Vafa, \emph{{Compactifications of F theory on Calabi-Yau
  threefolds. 1}},
  \href{https://doi.org/10.1016/0550-3213(96)00242-8}{\emph{Nucl.Phys.}
  {\bfseries B473} (1996) 74}
  [\href{https://arxiv.org/abs/hep-th/9602114}{{\ttfamily hep-th/9602114}}].

\bibitem{oai:arXiv.org:hep-th/9603161}
D.R.~Morrison and C.~Vafa, \emph{{Compactifications of F theory on Calabi-Yau
  threefolds. 2.}},
  \href{https://doi.org/10.1016/0550-3213(96)00369-0}{\emph{Nucl.Phys.}
  {\bfseries B476} (1996) 437}
  [\href{https://arxiv.org/abs/hep-th/9603161}{{\ttfamily hep-th/9603161}}].

\bibitem{oai:arXiv.org:1111.1232}
T.W.~Grimm and H.~Hayashi, \emph{{F-theory fluxes, Chirality and Chern-Simons
  theories}}, \href{https://doi.org/10.1007/JHEP03(2012)027}{\emph{JHEP}
  {\bfseries 1203} (2012) 027}
  [\href{https://arxiv.org/abs/1111.1232}{{\ttfamily 1111.1232}}].

\bibitem{Krause:2012yh}
S.~Krause, C.~Mayrhofer and T.~Weigand, \emph{{Gauge Fluxes in F-theory and
  Type IIB Orientifolds}},
  \href{https://doi.org/10.1007/JHEP08(2012)119}{\emph{JHEP} {\bfseries 08}
  (2012) 119} [\href{https://arxiv.org/abs/1202.3138}{{\ttfamily 1202.3138}}].

\bibitem{Braun:2013nqa}
V.~Braun, T.W.~Grimm and J.~Keitel, \emph{{Geometric Engineering in Toric
  F-Theory and GUTs with U(1) Gauge Factors}},
  \href{https://doi.org/10.1007/JHEP12(2013)069}{\emph{JHEP} {\bfseries 12}
  (2013) 069} [\href{https://arxiv.org/abs/1306.0577}{{\ttfamily 1306.0577}}].

\bibitem{Cvetic:2013uta}
M.~Cveti{\v c}, A.~Grassi, D.~Klevers and H.~Piragua, \emph{{Chiral
  Four-Dimensional F-Theory Compactifications With SU(5) and Multiple
  U(1)-Factors}}, \href{https://doi.org/10.1007/JHEP04(2014)010}{\emph{JHEP}
  {\bfseries 04} (2014) 010} [\href{https://arxiv.org/abs/1306.3987}{{\ttfamily
  1306.3987}}].

\bibitem{Cvetic:2015txa}
M.~Cveti{\v c}, D.~Klevers, D.K.M.~Pe{\~ n}a, P.-K.~Oehlmann and J.~Reuter,
  \emph{{Three-Family Particle Physics Models from Global F-theory
  Compactifications}},
  \href{https://doi.org/10.1007/JHEP08(2015)087}{\emph{JHEP} {\bfseries 08}
  (2015) 087} [\href{https://arxiv.org/abs/1503.02068}{{\ttfamily
  1503.02068}}].

\bibitem{Lin:2015qsa}
L.~Lin, C.~Mayrhofer, O.~Till and T.~Weigand, \emph{{Fluxes in F-theory
  Compactifications on Genus-One Fibrations}},
  \href{https://doi.org/10.1007/JHEP01(2016)098}{\emph{JHEP} {\bfseries 01}
  (2016) 098} [\href{https://arxiv.org/abs/1508.00162}{{\ttfamily
  1508.00162}}].

\bibitem{Lin:2016vus}
L.~Lin and T.~Weigand, \emph{{G4-flux and standard model vacua in F-theory}},
  \href{https://doi.org/10.1016/j.nuclphysb.2016.09.008}{\emph{Nucl. Phys.}
  {\bfseries B913} (2016) 209}
  [\href{https://arxiv.org/abs/1604.04292}{{\ttfamily 1604.04292}}].

\bibitem{Krause:2011xj}
S.~Krause, C.~Mayrhofer and T.~Weigand, \emph{{$G_4$ flux, chiral matter and
  singularity resolution in F-theory compactifications}},
  \href{https://doi.org/10.1016/j.nuclphysb.2011.12.013}{\emph{Nucl. Phys.}
  {\bfseries B858} (2012) 1} [\href{https://arxiv.org/abs/1109.3454}{{\ttfamily
  1109.3454}}].

\bibitem{Cvetic:2018ryq}
M.~Cvetič, L.~Lin, M.~Liu and P.-K.~Oehlmann, \emph{{An F-theory Realization
  of the Chiral MSSM with $\mathbb{Z}_2$-Parity}},
  \href{https://doi.org/10.1007/JHEP09(2018)089}{\emph{JHEP} {\bfseries 09}
  (2018) 089} [\href{https://arxiv.org/abs/1807.01320}{{\ttfamily
  1807.01320}}].

\bibitem{Cvetic:2019gnh}
M.~Cvetič, J.~Halverson, L.~Lin, M.~Liu and J.~Tian, \emph{{Quadrillion
  $F$-Theory Compactifications with the Exact Chiral Spectrum of the Standard
  Model}}, \href{https://doi.org/10.1103/PhysRevLett.123.101601}{\emph{Phys.
  Rev. Lett.} {\bfseries 123} (2019) 101601}
  [\href{https://arxiv.org/abs/1903.00009}{{\ttfamily 1903.00009}}].

\bibitem{Bies:2014sra}
M.~Bies, C.~Mayrhofer, C.~Pehle and T.~Weigand, \emph{{Chow groups, Deligne
  cohomology and massless matter in F-theory}},
  \href{https://arxiv.org/abs/1402.5144}{{\ttfamily 1402.5144}}.

\bibitem{Bies:2017fam}
M.~Bies, C.~Mayrhofer and T.~Weigand, \emph{{Gauge Backgrounds and Zero-Mode
  Counting in F-Theory}},
  \href{https://doi.org/10.1007/JHEP11(2017)081}{\emph{JHEP} {\bfseries 11}
  (2017) 081} [\href{https://arxiv.org/abs/1706.04616}{{\ttfamily
  1706.04616}}].

\bibitem{Bies:2018uzw}
M.~Bies, \emph{{Cohomologies of coherent sheaves and massless spectra in
  F-theory}}, Ph.D. thesis, Heidelberg U., 2018-02.
\newblock \href{https://arxiv.org/abs/1802.08860}{{\ttfamily 1802.08860}}.

\bibitem{Bies:2020gvf}
M.~Bies, M.~Cveti\v{c}, R.~Donagi, L.~Lin, M.~Liu and F.~Ruehle, \emph{{Machine
  Learning and Algebraic Approaches towards Complete Matter Spectra in 4d
  F-theory}}, \href{https://doi.org/10.1007/JHEP01(2021)196}{\emph{JHEP}
  {\bfseries 01} (2021) 196}
  [\href{https://arxiv.org/abs/2007.00009}{{\ttfamily 2007.00009}}].

\bibitem{Database}
M.~Bies, M.~Cveti{\v c}, R.~Donagi, L.~Lin, M.~Liu and F.~Ruehle,
  ``{Database}.'' \url{https://github.com/Learning-line-bundle-cohomology},
  2020.

\bibitem{ToricVarietiesProject}
{The~Toric~Varieties~project~authors}, ``{The $\mathtt{Toric Varieties}$
  project}.'' (\url{https://github.com/homalg-project/ToricVarieties_project}),
  2019--2021.

\bibitem{Brill1874}
Brill, \emph{{Ueber die algebraischen Functionen und ihre Anwendung in der
  Geometrie. (Zus. mit Noether)}}, {\emph{Mathematische Annalen} {\bfseries 7}
  (1874) 269}.

\bibitem{Eisenbud1996}
D.~Eisenbud, M.~Green and J.~Harris, \emph{{Cayley-Bacharach theorems and
  conjectures}},
  \href{https://doi.org/10.1090/s0273-0979-96-00666-0}{\emph{Bulletin of the
  American Mathematical Society} {\bfseries 33} (1996) 295}.

\bibitem{Watari:2016lft}
T.~Watari, \emph{{Vector-like pairs and Brill--Noether theory}},
  \href{https://doi.org/10.1016/j.physletb.2016.09.006}{\emph{Phys. Lett. B}
  {\bfseries 762} (2016) 145}
  [\href{https://arxiv.org/abs/1608.00248}{{\ttfamily 1608.00248}}].

\bibitem{bies2021root}
M.~Bies, M.~Cveti\v{c}, R.~Donagi, M.~Liu and M.~Ong, \emph{{Root Bundles and
  Towards Exact Matter Spectra of F-theory MSSMs}},
  \href{https://arxiv.org/abs/2102.10115}{{\ttfamily 2102.10115}}.

\bibitem{2004math......4078C}
C.~Lucia, C.~Cinzia and C.~Maurizio, \emph{{Moduli of roots of line bundles on
  curves}}, {\emph{Trans. Amer. Math. Soc. 359} (2007) }
  [\href{https://arxiv.org/abs/math/0404078}{{\ttfamily math/0404078}}].

\bibitem{Kreuzer:1998vb}
M.~Kreuzer and H.~Skarke, \emph{{Classification of reflexive polyhedra in
  three-dimensions}},
  \href{https://doi.org/10.4310/ATMP.1998.v2.n4.a5}{\emph{Adv. Theor. Math.
  Phys.} {\bfseries 2} (1998) 853}
  [\href{https://arxiv.org/abs/hep-th/9805190}{{\ttfamily hep-th/9805190}}].

\bibitem{Batyrev:1994hm}
V.V.~Batyrev, \emph{{Dual polyhedra and mirror symmetry for Calabi-Yau
  hypersurfaces in toric varieties}},
  \href{https://doi.org/10.1090/jag/666}{\emph{J. Alg. Geom.} {\bfseries 3}
  (1994) 493} [\href{https://arxiv.org/abs/alg-geom/9310003}{{\ttfamily
  alg-geom/9310003}}].

\bibitem{Perevalov:1997vw}
E.~Perevalov and H.~Skarke, \emph{{Enhanced gauged symmetry in type II and F
  theory compactifications: Dynkin diagrams from polyhedra}},
  \href{https://doi.org/10.1016/S0550-3213(97)00477-X}{\emph{Nucl. Phys. B}
  {\bfseries 505} (1997) 679}
  [\href{https://arxiv.org/abs/hep-th/9704129}{{\ttfamily hep-th/9704129}}].

\bibitem{cox1999mirror}
{Cox, D.A. and Katz, S.}, \emph{{Mirror Symmetry and Algebraic Geometry}},
  {Mathematical surveys and monographs}, {American Mathematical Society}
  ({1999}).

\bibitem{Rohsiepe:2004st}
F.~Rohsiepe, \emph{{Lattice polarized toric K3 surfaces}},
  \href{https://arxiv.org/abs/hep-th/0409290}{{\ttfamily hep-th/0409290}}.

\bibitem{Braun:2017nhi}
A.P.~Braun, C.~Long, L.~McAllister, M.~Stillman and B.~Sung, \emph{{The Hodge
  Numbers of Divisors of Calabi-Yau Threefold Hypersurfaces}},
  \href{https://arxiv.org/abs/1712.04946}{{\ttfamily 1712.04946}}.

\bibitem{Halverson:2016tve}
J.~Halverson and J.~Tian, \emph{{Cost of seven-brane gauge symmetry in a
  quadrillion F-theory compactifications}},
  \href{https://doi.org/10.1103/PhysRevD.95.026005}{\emph{Phys. Rev. D}
  {\bfseries 95} (2017) 026005}
  [\href{https://arxiv.org/abs/1610.08864}{{\ttfamily 1610.08864}}].

\bibitem{cox2011toric}
D.~Cox, J.~Little and H.~Schenck, \emph{{Toric Varieties}}, {Graduate studies
  in mathematics}, American Mathematical Soc. (2011),
  \href{https://doi.org/10.1090/gsm/124}{10.1090/gsm/124}.

\bibitem{Kreuzer:2006ax}
M.~Kreuzer, \emph{{Toric geometry and Calabi-Yau compactifications}},
  {\emph{Ukr. J. Phys.} {\bfseries 55} (2010) 613}
  [\href{https://arxiv.org/abs/hep-th/0612307}{{\ttfamily hep-th/0612307}}].

\end{thebibliography}\endgroup

\end{document}